\documentclass{elsarticle}

\usepackage{amsmath}
\usepackage{graphicx}
\usepackage{fancyhdr}
\usepackage[mathscr]{euscript}
\usepackage{amsmath}
\usepackage{amssymb}
\usepackage{amsthm}
\usepackage{algorithm2e}
\usepackage{epstopdf}
\usepackage{float}

\usepackage{amssymb}

\journal{arxiv}

\begin{document}

\begin{frontmatter}



\title{Robust Energy Storage Scheduling for Imbalance Reduction of Strategically Formed Energy Balancing Groups}


\author[NEC]{Shantanu Chakraborty\corref{cor1}}
\ead{s-chakraborty@cp.jp.nec.com}

\author[NEC]{Toshiya Okabe}

\cortext[cor1]{Corresponding author}

\address[NEC]{Smart Energy Research Laboratories, Central Research Laboratories,
NEC Corporation, Japan.}


\begin{abstract}
Imbalance (on-line energy gap between contracted supply and actual demand, and associated cost) reduction is going to be a crucial service for a Power Producer and Supplier (PPS) in the deregulated energy market. PPS requires forward market interactions to procure energy as precisely as possible in order to reduce imbalance energy. This paper presents, 1) (off-line) an effective demand aggregation based strategy for creating a number of balancing groups that leads to higher predictability of group-wise aggregated demand, 2) (on-line) a robust energy storage scheduling that minimizes the imbalance for a particular balancing group considering the demand prediction uncertainty. The group formation is performed by a \emph{Probabilistic Programming} approach using Bayesian Markov Chain Monte Carlo (MCMC) method after applied on the historical demand statistics. Apart from the group formation, the aggregation strategy (with the help of Bayesian Inference) also clears out the upper-limit of the required storage capacity for a formed group, fraction of which is to be utilized in on-line operation. For on-line operation, a robust energy storage scheduling method is proposed that minimizes expected imbalance energy and cost (a non-linear function of imbalance energy) while incorporating the demand uncertainty of a particular group. The proposed methods are applied on the real apartment buildings' demand data in Tokyo, Japan. Simulation results are presented to verify the effectiveness of the proposed methods.
\end{abstract}


\end{frontmatter}


\section{Introduction}
\label{intro}
The number of Power Producer and Supplier (PPS) in the power and energy market is increasing rapidly due to the liberalization in power market in Japan \citep{meti}. With the increase in market share of PPSs, the potential of demand-centric business opportunity increases. As of 2014, the electricity market in Japan is dominated by regional monopolies, where 85\% of the installed generating capacity is produced by 10 privately owned companies. However, the rising of Power Producer and Supplier (PPS) (i.e. Electric Power Retailer) in the electricity market is inevitable due to the full-fledged deregulation \cite{meti} that will eventually break the monopolies.
PPPs face challenge while keeping the on-line supply and demand matched with the highest precision, and thus reducing the imbalance (gap between contracted supply and demand) in demand side low-voltage network. In off-line, the PPS can intelligently group the customers to increase the demand predictability of each group and procures volume of energy (utilizing day-ahead energy prediction). The procured energy refers as the \emph{supply contract} for each group. Flexible power distribution as such is attainable through Digital grid architecture \citep{Abe:2011}. In ideal world, the contracted supply matches with actual demand at each granular (typically, 30-minutes). However, due to the uncertainty in on-line energy consumption as well as energy supply, the gap between supply and demand is highly likely to occur. The current practice is to buy (in case of demand is higher than the supply) or sell (in case of supply is higher than the demand) energy from/to Energy Imbalance Market (EIM) (EIM can be a part of Utility or be an independent body or the Utility itself, \citep{eim}). The EIM mitigates such mismatch between on-line supply and demand by transacting necessary energy with the PPS. The involvement of EIM goes higher with the increasing gap between the supply and the demand. The price setting of EIM, on the other hand, is significantly higher compared to the conventional energy tariff (\citep{eim}, page 4). Therefore, the reduction of imbalance cost casts itself as one of the important problems to tackle for demand side based energy service of PPS. 

Two fundamental yet interconnected problems are, therefore, identified for a PPS, 1) strategic demand aggregation for balancing group creation, and 2) on-line imbalance energy and cost reduction for formed balancing groups. A PPS serves multiple commercial settings (e.g. apartment buildings, commercial buildings, shopping mall, factory, etc.). Therefore, it is essential for the PPS to effectively and strategically identify the customers' demand based grouping for energy balancing purpose, i.e. \emph{balancing group}. Groupings as such are also necessary for service and price differentiations. In case of the imbalance energy reduction service, it is critically important for the PPS to define appropriate demand aggregation criterion and demand aggregation strategy so that it can effectively identify clusters of similar customers (e.g. buildings), the associated aggregated demand with reduced variance and potential imbalance energy bound. In this paper, we present a \emph{probabilistic programming} \citep{gordon2014probabilistic} approach that utilizes a Bayesian Markov Chain Monte Carlo (MCMC) sampling method \citep{geyer2011introduction} \citep{gelman2014bayesian} in order to form multiple balancing groups. Initially, a demand aggregation criterion is identified as a statistical \emph{observation}, then a probabilistic model of the \emph{observation} is devised and finally posteriors of the model parameters are determined through Bayesian MCMC. The process is recursively conducted, which divides a parent \emph{observation} into two based on the posterior analysis of the model parameters. Therefore, a divide-and-conquer approach is designed to solve balancing group formation problem. 

As for the on-line operation, the reduction of the group-wise imbalance cost can be realized by the effective on-line energy storage management; more particularly, the on-line charge/discharge (CD) scheduling of energy storage. 
However, since the imbalance energy tariff is non-linear to the imbalance energy, typical straight-forward method of CD scheduling leads to an inefficient solution. Therefore, we present an efficient CD scheduling for multiple spatially distributed energy storages where the schedule is robust against demand prediction uncertainty. The robust scheduling method essentially minimizes the expected imbalance energy and imbalance cost considering a number of demand prediction scenarios. Battery storage system is utilized as the energy storage system. The designed scheduling approach first performs a short-term demand prediction, then generates a number of statistical scenarios of the predicted demands utilizing a joint distribution of probability density function (PDF) of prediction error with the PDF of variability of preceding periods; and finally solves a multi-objective optimization problem that decides the CD scheduling (with power dispatch) of batteries while minimizing both imbalance energy and cost. The required aggregated battery power-rating information is drawn from the posterior distribution knowledge that had been conducted while forming groups. The problem in hand is non-linear due to the imbalance pricing scheme, storage dynamics and associated non-linear constraints. The optimization problem is, therefore, transformed into an equivalent Mixed Integer Linear Programming (MILP) problem \citep{Boyd:2004:CO:993483} followed by an additional transformation to Mixed Logical Dynamical (MLD) System \citep{Bemporad1999407}, and finally solved by a branch-and-cut linear solver \citep{cbc}.

\subsection{Related Works}
\emph{Clustering} has been a useful analytical and operational tool in energy domain (e.g. energy market, \citep{Dias20147722}). Applications of clustering methods in high- and medium-voltage power networks for large scale integration of customers have been reported in articles like \citep{Chicco201268} and \citep{1626400}. In \citep{Chicco201268}, a comprehensive overview of clustering methods are presented and the necessity of these methods while identifying effective customer grouping are highlighted (from the perspective of an Energy Supplier). The supervised clustering algorithms (e.g. Hierarchical clustering, K-means, Fuzzy K-means) are discussed \cite{1626400} that analyzes the similarity within customers. Meanwhile, at the low-voltage network, energy based clustering for planning and operation (from the perspective of market operators such as Distribution Network Operators, DNOs) is reported in \citep{7063233}. In \citep{7063233}, household smart-meter data are analyzed for demand variability and a finite mixture model based clustering algorithm is proposed that discovers a number of distinctive behavior groups. Therefore, it is evident that clustering methods play important roles for planning and operation of market operators such as energy suppliers, DNOs and PPSs. We employed a Bayesian inference coupled with MCMC method to determine the energy balancing groups based on a statistical demand measurement. The Bayesian MCMC is chosen over other clustering methods due to its advantages of accounting the uncertainty presented in the models and parameters as well as its ability to present useful insights regarding the model (inferred from the posterior distribution of model's parameters). For example, the applied Bayesian MCMC method provides an upper-bound of required group-wise energy storage aggregated power-rating (and consequently energy capacity rating).

For on-line operation, which requires fast solution, it has been widely advised by the experts to deploy advanced mathematical optimization algorithms that accounts for system uncertainty and predictability of future states or conditions \citep{firestone2005energy}. 
The working mechanism of the proposed CD scheduling method, therefore, aligns with that of Model Predictive Control (MPC), which measures up to the aforementioned requirements. MPC (and its variants) has been an active research area in the power system arena. In \citep{6417004}, an energy management system for Microgrid operation considering PV, Diesel Generators, and energy storage is presented. A stochastic MPC for solving Unit Commitment with wind power is documented in \citep{5587912}. Robust scheduling of resources is critically important when the optimization model is exposed to various uncertainties presented in the model states. For example, in \citep{Wang2015597}, a robust cost optimization method is presented that essentially schedules of renewable energy generators with combined heat and power (CHP) generators considering the uncertainty in net energy demand and electricity price. Another stream of research has been conducted on stochastic MPC considering the uncertainty in the model. For example, in \citep{6802411}, a stochastic MPC is presented for efficient controlling in building's HVAC system while focusing on energy minimization. On the other hand, MILP based mathematical optimization is the current industry trend of operation research oriented towards resource scheduling and optimization. For instance, in the arena of Unit Commitment (one of the important problems in Power System Planning and Operation), MILP provides efficient, fast and scalable solutions \cite{viana2013new} \cite{1664974}. These outstanding researches create platform for applying robust MILP based optimization algorithm, which minimizes the expected imbalance energy and cost over a set predicted demand scenarios, as a core optimizer of the proposed on-line CD scheduling method.

\section{Forming Energy Balancing Groups}
\label{sec_energy_bal}
An energy balancing group contains a number of customers\footnote{Being specific to this manuscript, the customers are the apartments buildings (or simply buildings).} sharing similarity in their demand profiles where their aggregated demand profile exhibits higher predictability. It is, therefore, essential to identify appropriate criterion based on which such aggregation and group formation will be performed. The following is one such criterion.

\subsubsection{Maximum Demand Standard Deviation of Periodic Demand}
This measure is defined as the maximum of demand standard deviation ($DSD$), $MDSD$ of a particular customer. The $DSD$ tells how the demand of a particular period deviates from the average demand of that period over a range of sampling days. For example, if the demand is sampled over January (containing 31 days), the $DSD$ of 10 AM is the statistical standard deviation over all the demands at 10AM accumulated in January. The $MDSD$ is thus defined as the following
\begin{equation}
MDSD = max\left\{\sigma_{t}(D_{i}(t))\right\}.
\label{eqn_mdsd}
\end{equation}
where $t=1,...,N$ ($N$ is the number of periods in a day, e.g. $N=48$ for a day of 30-minutes granularity) and $i=1,...,ND$ ($ND$ is the total number of sampling days).
$MDSD$ is an absolute (demand) measure that stems from the in-period demand variations over different days.

From a PPS's perspective, the customers that lower the accumulated $MDSD$ are desirable, since those customers have similarities in demand pattern as well as higher predictability in their aggregated demand\footnote{So called \emph{Laws of Large Numbers}.}. $MDSD_{c}$ is, therefore, identified as the demand aggregation criterion $DAC_{c}$, for a particular customer $c$. $MDSD$ is considered to be an extremely important criterion since it can provide insights regarding potential imbalance that might occur in real-time. More specifically, $MDSD$ says about the worst deviation (from the average demand) that likely to occur over a particular day, consequently, the maximum power required to nullify the deviation. As batteries are deployed to countermeasure the imbalance, certain insights regarding the required battery power is essential. Therefore, $MDSD$ seems to be a better choice over other criteria as far as imbalance reduction is concerned. Other criteria, such as medians of daily load factor, total periodical demand, daily average demand, etc. could also be utilized (either stand-alone or in combination) as $DAC_{c}$ depending on the goal of the design. In this paper, therefore, we limit the $DAC$ only to $MDSD$.

\subsection{Demand Aggregation Strategy}
The demand aggregation strategy performs a demand grouping scheme that provides several effective balancing groups of customers. The strategy first determines the $DAC$s for all customers using their historical demand profiles. The strategy then applies a \emph{divide and conquer} algorithm that recursively utilizes a \emph{Probabilistic Programming} on the DACs. The Bayesian MCMC sampling method is chosen as the \emph{Probabilistic Programming}. The \emph{Probabilistic Programming} approach basically designs a probabilistic model for observing the $DAC$s (referred as \emph{observation}) by utilizing statistical knowledge regarding model parameters (such as, when and how the observation changes with the arrival of new customer), represented by associated probability distributions and later utilizes Bayesian inference to generate posterior distributions of the model parameters (by sampling through MCMC).

\begin{figure}[htb]
\centering
\includegraphics[scale=0.4]{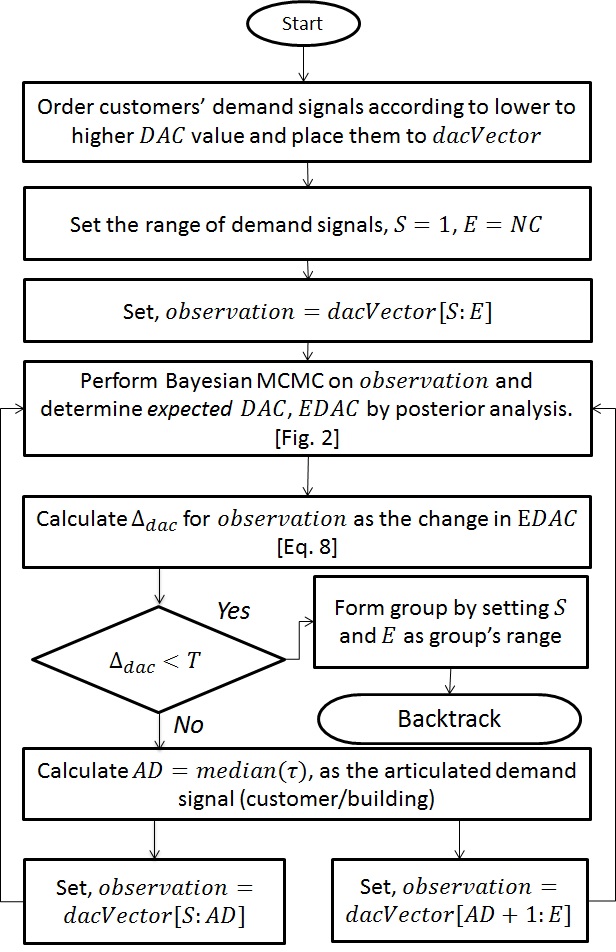}
\caption{Flow-chart of a divide and conquer based group formation (demand aggregation strategy).}
\label{fig_demand_strategy}
\end{figure}

Figure~\ref{fig_demand_strategy}: \textit{Flow-chart of a divide and conquer based group formation (demand aggregation strategy).}

\begin{figure}[htb]
\centering
\includegraphics[scale=0.5]{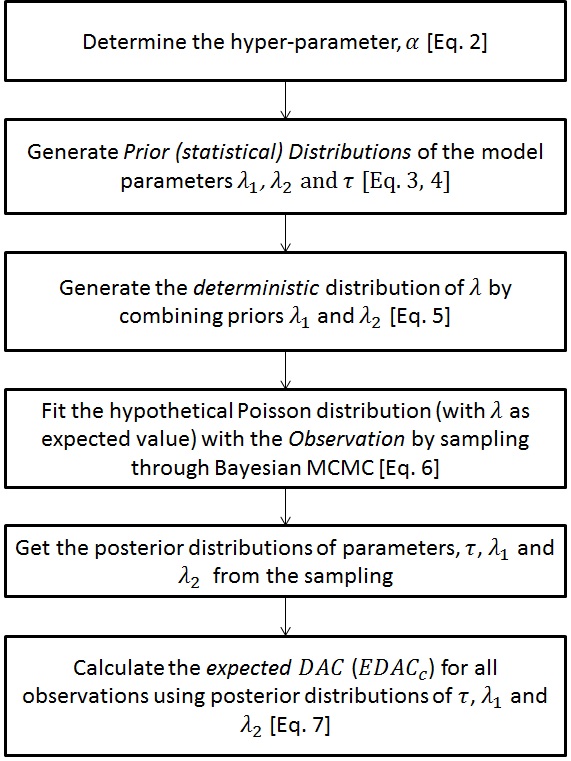}
\caption{Flow-chart of the Bayesian MCMC sampling method generating posterior distributions of the model parameters.}
\label{fig_mcmc}
\end{figure}

Figure~\ref{fig_mcmc}: \textit{Flow-chart of the Bayesian MCMC sampling method generating posterior distributions of the model parameters.}

The process of Demand Aggregation Strategy (i.e. demand based energy balancing group formation) is depicted in Figure~\ref{fig_demand_strategy}. 
In the flowchart, $NC$ is number of individual customers. Note that, the method divides a group into two and recursively solves the sub-group formation problem. The output of the process is a list of group information where each item in the list represents the range of the group (start and end). In detail, as depicted in Fig~\ref{fig_demand_strategy}, the customers are organized in ascending order according to their DACs. The initial balancing group contains all of the customers (i.e. a single group). The initial observation contains the DAC values of all customers. The next step is to build the probabilistic model and use Bayesian MCMC sampling method to generate posterior distributions of model parameters. Fig~\ref{fig_mcmc} shows the flow-chart of the group partition process. Basically, the \emph{observation} is where the model trying to fit-in by varying the model parameters. At first the hyper-parameter (the parameter that controls the other parameters) $\alpha$ is determined. The parameter $\alpha$ (Eq. \ref{alpha}, where $NO$ is number of observations) is set as the inverse of the expected observations and is used to parameterize the prior distributions of $\lambda_{1}$ and $\lambda_{2}$ (Eq. \ref{lambda12}). What follows are the key features of the process.
\begin{itemize}
\item[1.] Uniform distribution for the \emph{articulated customers}\footnote{The term ``articulated customers'' is defined as the customers whose DAC differs `significantly' with the same of preceding customers when the customers are ordered ascending-ly according to their DACs. In other words, an \emph{articulated customer} defines the boundary of potential energy balancing groups. Every customer is equally likely to be an articulated customer (prior belief; before an observation of DAC is made). Therefore, the probability (parameter $\tau$) of a particular customer to be an articulated customer is uniformly distributed over the number of customers}; parameter, $\tau$ (Eq. \ref{tau}).
\item[2.] \emph{Exponential Distributions} of DAC before and after the \emph{articulated customers}; parameters, $\lambda_{1}$ and $\lambda_{2}$; which in turn parameterized by $\alpha$ (Eq \ref{lambda12}).
\item[3.] $\lambda$ is formed deterministically by combining $\lambda_{1}$ and $\lambda_{2}$ where the merging point is determined by distribution $\tau$ and customer identifier $c$ (Eq. \ref{lambda}).
\item[4.] The $DAC$ values are hypothesized by a \emph{Poisson Distribution}\footnote{The distribution of $DAC$ is chosen to be a \emph{Poisson Distribution}, since the $DAC$ of a customer is independent of each other and can be treated as a form of \textit{count} data occurred in a discrete time event.} with $\lambda$ as expected value (Eq. \ref{dac}).
\end{itemize}

\begin{figure}[htb]
\centering
\includegraphics[scale=0.6]{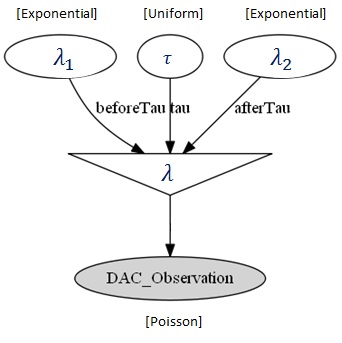}
\caption{Model parameters and internal dependencies.}
\label{fig_stochastic_model}
\end{figure}

Figure~\ref{fig_stochastic_model}: \textit{Model parameters and internal dependencies.}

The probabilistic model with parameters and dependencies is plotted in Figure~\ref{fig_stochastic_model}. The following equations mathematically show the definitions of distributions and associated parameters.
\begin{equation}
\alpha=\left [ \frac{1}{NO}\sum_{c=S}^{E}DAC_{c} \right ]^{-1}
\label{alpha}
\end{equation}

\begin{eqnarray}
\label{lambda12}
\lambda_{1}\sim Exp(\alpha) \\ \nonumber 
\lambda_{2}\sim Exp(\alpha)
\end{eqnarray}

\begin{equation}
\tau \sim DiscreteUniform(1, NO)
\label{tau}
\end{equation}

\begin{equation}
\lambda=\left\{\begin{matrix}
\lambda_{1} & if~c\leq \tau \\ 
\lambda_{2} & if~c> \tau
\end{matrix}\right.
\label{lambda}
\end{equation}

\begin{equation}
DAC_{c} \sim Poisson(\lambda)
\label{dac}
\end{equation}

\begin{equation}
EDAC_{c}=\frac{1}{NO}\left ( \sum_{\forall \tau_{i}>c}\lambda_{1,i} + \sum_{\forall \tau_{j}\leq c}\lambda_{2,j} \right )
\label{edac}
\end{equation}

\begin{equation}
\Delta_{dac} = \beta \times \frac{EDAC_{E}}{EDAC_{S}}
\label{ddac}
\end{equation}
Utilizing Bayesian MCMC method, the created probabilistic model is conditioned to fit the observed $DAC$ into a \emph{Poisson Distribution}. The posterior distributions of the model parameters ($\tau$ and $\lambda$s) are resulted from the method. The process (Fig~\ref{fig_mcmc}), after determining the posterior distributions of the statistical parameters, calculates the expected $DAC$ ($EDAC$), according to Eq. \ref{edac}. The group partition criteria is then determined by the change in $DAC$ ($\Delta_{DAC}$, Eq. \ref{ddac}). The group formation method (Fig~\ref{fig_demand_strategy}) utilizes $\Delta_{DAC}$ (with a threshold $T$) to divide a single group into two and recursively works on each of the groups to identify a number of energy balancing groups. The MCMC process utilizes the Metropolis-Hastings algorithm \cite{geyer2011introduction} \cite{gelman2014bayesian} while fitting the probabilistic model to the target distribution (i.e. \emph{observation}). The details of the algorithm and underlying theory can be found at \cite{gelman2014bayesian}.

\section{On-line Storage Scheduling for Imbalance Reduction}
\begin{figure}[htb]
\centering
\includegraphics[scale=0.5]{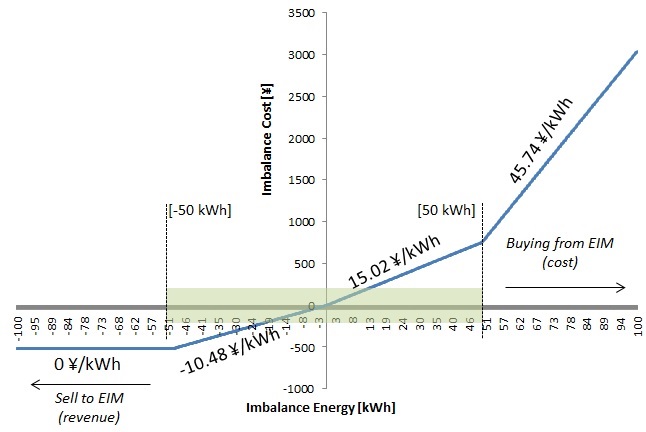}
\caption{Non-linear convex imbalance pricing scheme.}
\label{fig_imbalance_pricing}
\end{figure}

In this section, an on-line stochastic scenario based robust storage scheduling method is proposed that reduces the imbalance energy and imbalance cost of a particular balancing group. Even, the balancing groups are formed strategically in order to maximize the aggregated demand predictability (which is utilized in the forward and day-ahead market to determine supply contract), in real-time, the imbalance of energy is still inevitable due to the prediction error. The PPS, therefore, interacts with the Energy Imbalance Market (EIM) to nullify the imbalance (by purchasing in case of demand is higher than the contracted supply or by selling otherwise). The price setting of EIM is considerably higher compared with conventional energy tariff. Exemplary imbalance pricing scheme is shown in Figure~\ref{fig_imbalance_pricing} \citep{eim}. The pricing scheme follows a nonlinear curve where a higher penalty has to be paid by PPS if the imbalance energy goes beyond a threshold (in Fig.~\ref{fig_imbalance_pricing} the threshold is set as 50kWh), in case of buying from EIM. On the other hand, PPS will receive no additional revenue if the energy to be sold is higher than the threshold (-50kWh).
The threshold is set as 3\% of the monthly peak supply, according to \citep{eim}.

Figure~\ref{fig_imbalance_pricing}: \textit{Non-linear convex imbalance pricing scheme.}

\begin{figure}[htb]
\centering
\includegraphics[scale=0.5]{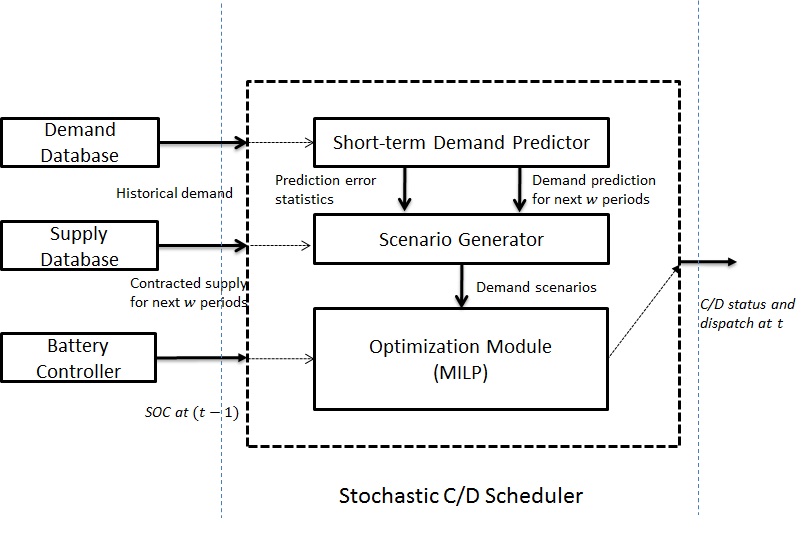}
\caption{Outline of S-SWCD algorithm with components.}
\label{fig_cd_scheduler}
\end{figure}

Therefore, reducing imbalance energy and cost are narrowed down to optimizing the non-linear convex imbalance pricing curve by controlling storage charge/discharge power. To this end, we propose a stochastic sliding window based charge/discharge (CD) algorithm (that is designed based on the concept of MPC), which will reduce the imbalance energy (with imbalance cost) by intelligently charging/discharging spatially distributed energy storages (e.g. battery). We refer the method as Stochastic Sliding Window based CD algorithm (S-SWCD). The system outline for S-SWCD is described in Figure~\ref{fig_cd_scheduler}. The system receives (at a particular period $t$) contracted supply for a short window (let's say for next $w$ periods), aggregated demand of recent past, and current measured status of the batteries (state-of-charge, SOC) and produces an optimal CD schedule for a number of batteries that eventually reduce the imbalance energy and imbalance cost and are robust against the demand uncertainty. The Optimization Module consists of an MILP solver that is responsible for producing robust CD scheduling after minimizing the expected imbalance energy and imbalance cost. Although, the module produces CD scheduling for the next $w$ periods, only the 1st CD schedule (which is for period $t$) is applied to the battery system and the rest of the schedules are discarded. S-SWCD provides a closed loop solution since a feedback policy is implemented to compensate the demand variability and uncertainty. The subsequent sections will describe the components of the system. 

Figure~\ref{fig_cd_scheduler}: \textit{Outline of S-SWCD algorithm with components.}

\subsection{Short-term Demand Predictor}
A Support Vector Machine (SVM) \cite{burges1998tutorial} based time series prediction methodology is applied in order to predict the demand by utilizing the historic demand information. The demand signal is a time series, which follows a certain trend line (regulated by e.g. periods, weekdays, holidays, etc.). SVM finds optimal regression (Support Vector Regression, SVR) models while minimizing the training error and model complexity. The developed SVR based demand prediction engine models the (recent) past demand patterns to predict demand for a short window (for a window size of $w$, typically for next 4-hours). At a certain period $t$, the predictor produces the estimated demand $\widetilde{Dm}_{t}$ to $\widetilde{Dm}_{t+w-1}$ using the historical demand till $t-1$. This expression can be written as $\widetilde{Dm}_{t+i|t-1}$, $i=0,...,w-1$. The predictor creates $w$ separate models for each of the lagged periods. For example, while predicting demand $\widetilde{Dm}_{t|t-1}$, SVR engine creates and trains the model (of lag 1) by fitting a particular demand $Dm_{i}$ with a non-linear mapping of its previous demands, starting from demand at $t-1$ down to demand at a particular training horizon. Note that, the training data set only considers the recent demand set (instead of the whole data set) to avoid over-training the model by seasonally differed demand data. SVR tries to generate the non-linear model as the following function $f_{model}^{l}$ for lag $l$ ($l=0,...,w-1)$,
\begin{equation}
f_{model}^{l}:\left [Dm_{i-l-1},Dm_{i-l-2},...,Dm_{i-NP-1},F_{i} \right ]\mapsto Dm_{i}
\end{equation}
where $i=t-TH-1,...,t-1$, $TH$ is the training horizon, $NP$ is the number of past periods, $F_{i}$ is additional feature vector containing influential temporal information such as, holiday/weekend indicator, time of the day, and day of the week. The radial basis function is used as SVM kernel that transforms the data into a higher dimensional space while performing the regression. The hyper-parameters for creating the appropriate model are fixed by performing appropriate number of cross-validations within fractions of training data (so-called through grid-search).

However, the predicted demand for a particular period tends to change due to uncertainty. For example, $\widetilde{Dm}_{t+2|t}$ is not necessarily similar to $\widetilde{Dm}_{t+2|t+1}$. That is, the predicted demand at 12:00 when predicting at 11:00 is not necessarily same as when predicting at 11:30. Which is why, a deterministic objective function is not capable of handling the uncertainty imposed by the demand predictor. Therefore, a stochastic scenario based optimization approach is undertaken that minimizes the expected imbalance energy and imbalance cost. The next section focuses on the Scenario Generator.

\subsection{Scenario Generator}

\begin{figure}[htb]
\centering
\includegraphics[scale=0.45]{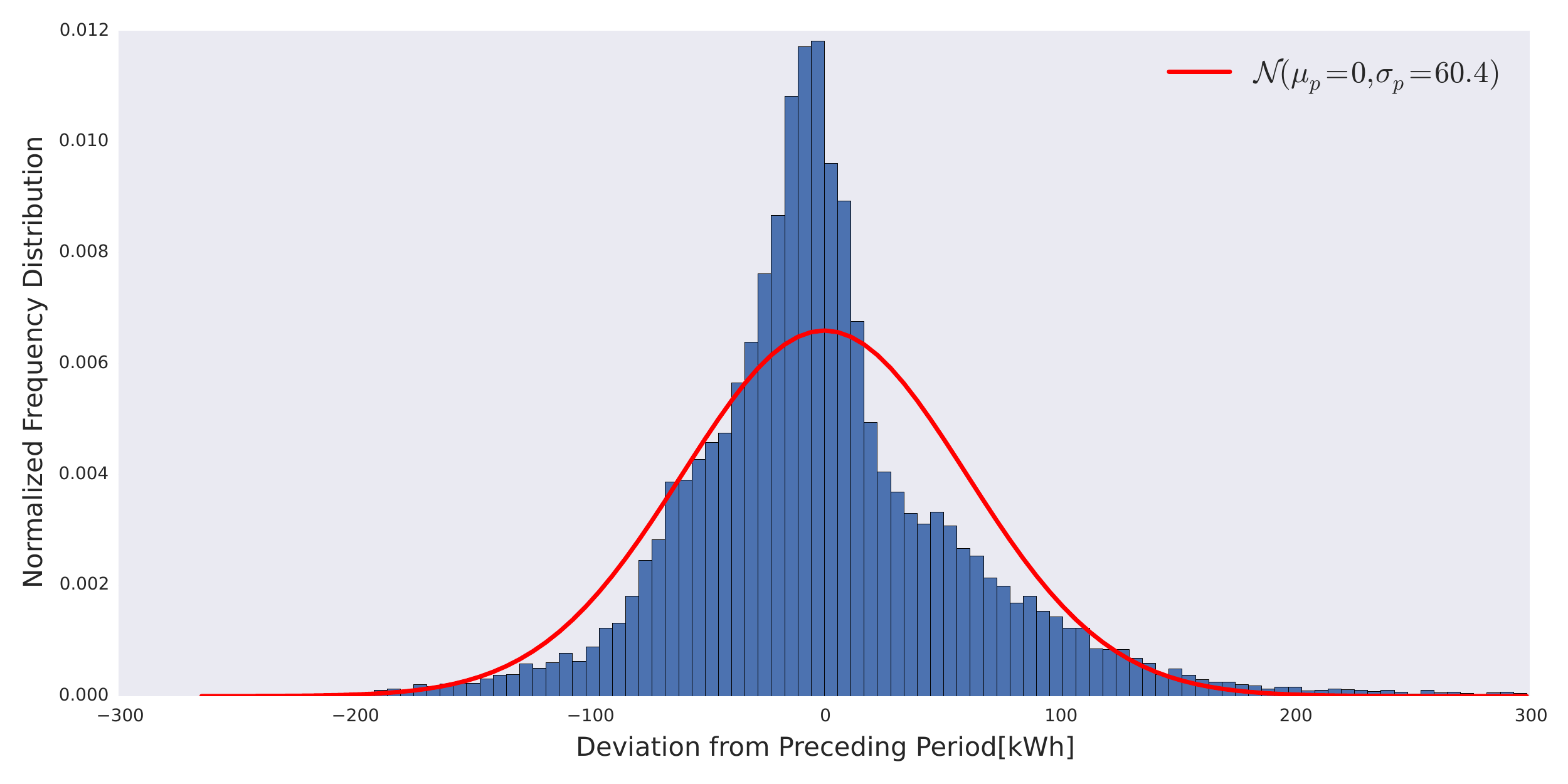}
\caption{Distribution of \emph{demand variability} from preceding period with PDF.}
\label{fig_pdf_preced}
\end{figure}

A \emph{scenario}, in this context, is actually a randomized snapshot of a predicted demand signal. The scenario is generated by utilizing the demand prediction error information coupled with the variability from the demand of preceding period. The system gradually learns the prediction errors (that are realized so far) for each of the lagged periods, $l$ and determines the \emph{probability density function}, PDF of the prediction error. The PDF of the prediction errors is a \emph{Gaussian Distribution} with almost zero mean and a specific standard deviation (let's call it the \emph{error PDF}; $\mathcal{N}(\mu_l, \sigma_l)$). On the other hand, the PDF of the \emph{demand variability} from the preceding period follows the Gaussian Distribution as well, as shown in Figure~\ref{fig_pdf_preced} (let's call it the \emph{variability PDF; $\mathcal{N}(\mu_{p}, \sigma_{p})$}). Therefore, the predicted demand scenarios are generated by collecting samples in a \emph{Bivariate Gaussian Distribution} of \emph{error PDF} and \emph{variability PDF}.
There are several ways to generate statistical scenarios, such as utilizing nonlinear programming for multi-stage decision problem \citep{hoyland2001generating}). However, sophisticated method as as requires higher computational power which goes against the speed requirement for an on-line operation. Therefore, we adopted a simpler yet effective ways to generate scenarios. Initially, a large \emph{scenario space} is generated utilizing the aforementioned \emph{Bivariate Gaussian Distribution}. The distribution is shown in the following equation
\begin{equation}
d_{l}(s\in \mathcal{S}) \sim \mathcal{N}(\mu,\Sigma)
\end{equation}
where $\mu$ contains the 2D-vector of means containing the $\mu_{l}$ and $\mu_{p}$, and $\Sigma$ is the covariance matrix containing the variances, shown as

\begin{equation}
\Sigma = \left [ \begin{matrix} 
\sigma_l^2 & \sigma_p^2  \\
\sigma_p^2  & \sigma_l^2 
\end{matrix}\right ]
\end{equation}
Therefore, the predicted demand for scenario $s\in \mathcal{S}$ is determined as following
\begin{equation}
\widetilde{Dm}_{t+l|t-1,s}=\widetilde{Dm}_{t+l|t-1} + d_{l}(s)
\label{eqn_scn_gen}
\end{equation}

\begin{figure}[htb]
\centering
\includegraphics[scale=0.45]{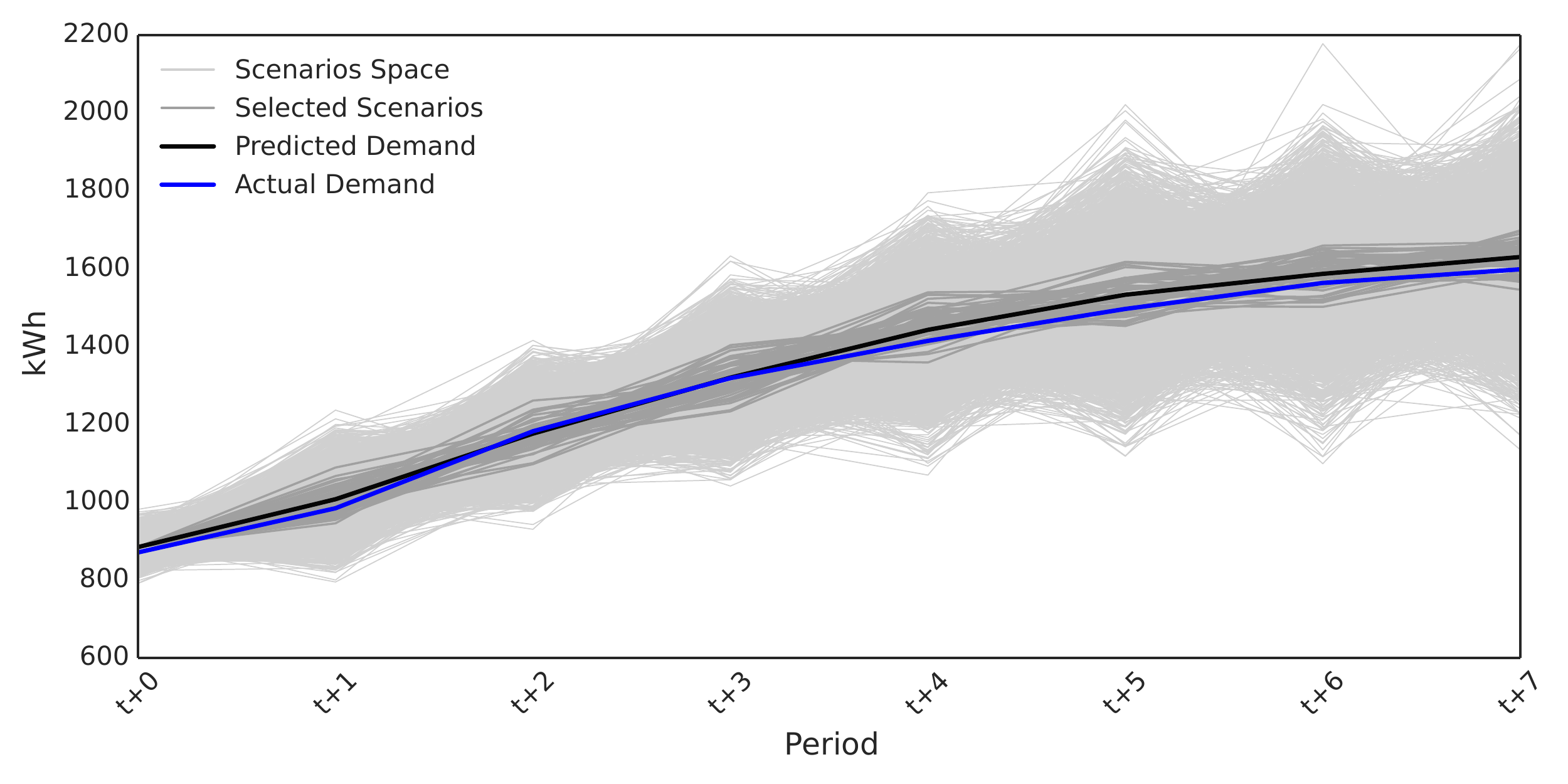}
\caption{Scenario generation (of 57 scenarios out of a scenario space of 5000 scenarios) for window size 8.}
\label{fig_scn_gen}
\end{figure}

Figure~\ref{fig_pdf_preced}: \textit{Distribution of \emph{demand variability} from preceding period with PDF.}

where $\widetilde{Dm}_{t+l|t-1,s}$ is the predicted demand scenario for lag $l$ and $\widetilde{Dm}_{t+l|t-1}$ is the predicted demand (from Demand Predictor module) at period $t+l$, predicted at period $t-1$. However, the \emph{scenario space} too large to be integrated into the optimization. Therefore, the space must be reduced (the so-called \emph{scenario reduction}). The reduction process is conducted by taking a subset of the \emph{scenario space}, $S\subset \mathcal{S}$. The elements $\widetilde{Dm}_{t+l|t-1,s}$ for $s\in S$ are chosen by considering the \emph{sum-of-squared distance} from the baseline predicted demand signals, $\widetilde{Dm}_{t+l|t-1}$.
 Figure~\ref{fig_scn_gen} shows a case of scenario generation for a particular predicted demand signal (considering $w=8$, number of scenarios = 57 and a 30-minutes granularity) with associated \emph{scenario space}.

Figure~\ref{fig_scn_gen}: \textit{Scenario generation (of 57 scenarios out of a scenario space of 5000 scenarios) for window size 8.}

\subsection{Optimization Module}
The optimization module reduces the imbalance energy and imbalance cost for a particular window (utilizing the predicted demand signals and scenarios) while deciding the storage charge/discharge schedule and associated power dispatch. The module however, utilizes the decision regarding CD schedule and dispatch for the current period while discarding the rest. At the next cycle, the module slides the window one time step and repeats the process (so called, closed-loop system).

\subsubsection{Objective Function with Constraints}
The problem is a multi-objective (due to imbalance energy and imbalance cost minimization) and stochastic optimization problem. The imbalance energy and imbalance cost need to be optimized separately due to non-linearity in the cost function (Figure~\ref{fig_imbalance_pricing}). The scalarized version of the stochastic optimization problem is described in Eq. \ref{eq_obj}. The 1st part of the objective function describes the absolute reduction of imbalance energy (weighted by $c_{0}$) while the 2nd part is about reduction of associated imbalance cost (weighted by $c_{1}$). Utilizing only imbalance cost as the objective will make the PPS intentionally increase the imbalance energy (by deliberately charging the battery up) so that it can sell the energy in later time, thereby, staying at the left side of the imbalance pricing curve. By incorporating the imbalance energy separately and performing the multi-objective optimization (with a small scaler weight given to the imbalance energy reduction), the system sustains a certain regulations on the energy and price trade-off. However, the system can always perform a single objective (only the imbalance cost, for instance) if it is the expected behavior by setting the corresponding weight as zero.

\begin{equation}
\label{eq_obj}
\begin{aligned}
& \underset{X_{i,b},P_{i,b}}{\text{minimize}} & & \mathbb{E}[FC(X,P)] \\
& & & :=\sum_{s\in S}Pr(s)\times \left [ 
\begin{split}
& c_{0}\times \sum_{i=t}^{t+w-1}\left | Im_{i,s} \right | + \\
& c_{1}\times\sum_{i=t}^{t+w-1}IC(Im_{i,s})
\end{split}
 \right ]
\end{aligned}
\end{equation}
where $FC$ is the cost function comprising the weighted imbalance energy and cost, $S$ is the predicted demand scenario set and $Pr(s)$ is the probability assigned to scenario $s$ (is set as $1/|S|$). The imbalance energy for a scenario $s\in S$ is formulated as (Eq. \ref{eq_const_im}).
\begin{equation}
Im_{i,s}= Sp_{i}-\widetilde{Dm}_{i,s}-\sum_{b\in B}P_{i,b}
\label{eq_const_im}
\end{equation}
where $Sp_{i}$ is the contracted supply at period $i$ and $\widetilde{Dm}_{i,s}$ is the predicted demand at period $i$ for scenario $s$. The imbalance cost function, $IC$ is a convex and non-linear function that can be presented by Figure~\ref{fig_imbalance_pricing}. The $IC$ contains both \emph{cost} part (when demand is higher than the supply) and \emph{revenue} part (when supply is higher than the demand).
$P_{i,b}$ is the power dispatch (+ve for charging, -ve for discharging) from/to a battery $b$ over a set of batteries $B$. The storage dynamics are presented as the following constraint.
\begin{eqnarray}
X_{i,b}=X_{i-1,b}+ \eta_{b} \times P_{i,b} \\
X_{b,min} \leq X_{i,b} \leq X_{b,max} \\
-p_{d,b}\leq P_{i,b} \leq p_{c,b}
\end{eqnarray}
Eq. 13 is the discrete time energy status (state-of-charge, SOC) of $b$ at period $i$ (considering $\Delta t=1$\footnote{thereby making $P_{i,b}$ as energy dispatch at period $i$ (done for simplification)}), where the efficiency $\eta_{b}$ is composed of charging efficiency $\eta_{b}^{c}$ and discharging efficiency $\eta_{b}^{d}$ (as shown in Eq. \ref{eq_cd_efficiency}).
\begin{equation}
\eta_{b}=\left \{\begin{matrix}\eta_{b}^{c},&if~P_{i,b}\geq 0\\ 1/\eta_{b}^{d},&otherwise \end{matrix} \right .
\label{eq_cd_efficiency}
\end{equation}
Eq. 14 bounds the SOC within a particular limit while Eq. 15 shows the power charging/discharging limit ($p_{c,b}$ and $p_{d,b}$, respectively). Due to the non-linearity in objective function (Eq. \ref{eq_obj}) and constraints (e.g. conditional CD efficiency at Eq. \ref{eq_cd_efficiency}), the optimization problem needs to be (equivalently) transformed into an MILP problem. The section to come describes the equivalent MILP formation of the above optimization problem.

\subsubsection{MILP Transformation}
In this section, the transformation of objective function (Eq. \ref{eq_obj}) and constraints to facilitate MILP formulations are presented.
The storage dynamics appeared in Eq. (13-15) can be effectively transformed into a linear formulation by introducing the following additional variables 
\begin{eqnarray}
S_{i,b}=\left \{ \begin{matrix} 1 & if~P_{i,b} \geq 0 \\ 0 & otherwise\end{matrix} \right . \\
Ax_{i,b}=S_{i,b}\times P_{i,b}
\end{eqnarray}
Therefore, utilizing MLD system formulation of converting logical dynamics to MILP \citep{Bemporad1999407}, the storage SOC dynamics can be transformed to 
\begin{equation}
X_{i,b}=X_{i-1,b}+(\eta_{b}^{c}-1/\eta_{b}^{d})\times Ax_{i,b} - 1/\eta_{b}^{d}\times Pb_{i,b}
\end{equation}
The non-linear logical constraints in Eq. (17-18) are equivalently casted to linear constraints \cite{Boyd:2004:CO:993483} and handled together with charging/discharging power limit of storage (Eq. 15). The followings shows the casted linear equations.
\begin{eqnarray}
 p_{d,b}\times S_{i,b} - P_{i,b} - p_{d,b} \leq 0 \\
-p_{d,b}\times S_{i,b} + P_{i,b}           \leq 0 \\
p_{d,b}\times S_{i,b} + Ax_{i,b} - P_{i,b} - p_{d,b} \leq 0 \\
p_{d,b}\times S_{i,b} - Ax_{i,b} + P_{i,b} - p_{d,b} \leq 0 \\
-p_{c,b}\times S_{i,b}+ Ax_{i,b} \leq 0 \\
-p_{c,b}\times S_{i,b}- Ax_{i,b} \leq 0
\end{eqnarray}
The non-linear and convex cost function requires to be linearized to be fitted into the MILP formulation. The transformation is conducted by introducing additional mixed-integer variables. For the sake of simplicity, we remove the scenario $s$ notation of original equations. The imbalance cost is equivalently transformed into a segmented combination of sub-costs (for a particular period $i$, as shown in below)
\begin{equation}
IC_{i} = IC(Im_{i}b) = \sum_{k=1}^{NPS}PS_{k}\times Z_{i,k}
\end{equation}
where $PS_{k}$, $k=1,...,NPS$ is the $k$-th imbalance unit price and $NPS$ is the number of price segments\footnote{As of Figure~\ref{fig_imbalance_pricing}, the pricing scheme has 4 unit price segments activated by an energy threshold (i.e. $\pm50$ kWh). There, $PS_{1}=45.7$, $PS_{2}=15.0$, $PS_{3}=10.48$, and $PS_{4}=0$ (JPY/kWh)\citep{eim}.}. The imbalance energy is constrained to be the sum of segmented energies, i.e.
\begin{equation}
Im_{i} = \sum_{k=1}^{NPS}Z_{i,k}
\end{equation}
The activation of $Z_{i,k}$ is controlled by binary variables $Y$. Considering $Th$ (as imbalance energy threshold, e.g. 50 kWh as of Figure~\ref{fig_imbalance_pricing}), a big number $M=1e+7$ and $NPS=4$; the following constraints are added to the formulation as a measure of activating appropriate $Z_{i,k}$ (avoiding the subscript $i$)
\begin{eqnarray}
(Th-M)\times Y_1 \leq Z_{1} \leq (Th-M)\times Y_0 \nonumber \\
-Th \leq Z_{2}\nonumber \\ 
Th\times Y_3 \leq Z_{3} \leq Th \times Y_2 \nonumber \\
0 \leq Z_{4} \leq M \times Y_3 \nonumber
\end{eqnarray}
Note that, the above equations can be generalized to work with any value of $NPS$. Some additional constraints are required to limit the activation in one of the halves of the pricing curve. 
The absolute value in the imbalance energy in Eq. \ref{eq_obj} is transformed into an equivalent function by introducing the lower- and upper-bound variables (since \emph{absolute value} imposes non-linearity and cannot be readily solvable by MILP).

\section{Numerical Simulations and Discussions}
This section presents the numerical simulations, analysis, results and discussions associated with formation of balancing groups and robust CD scheduling. A total of 103 apartments building in Tokyo are taken as customers and their demand data are utilized for the analysis. More particularly, the demand data of January and February, 2013 are taken and broken down to two phases
\begin{itemize}
\item[1.] Analysis and training phase: Data from January 1st to January 20th are utilized to perform the balancing group formation and train the initial prediction (SVR) model.
\item[2.] Optimization and simulation phase: Data from January 21st to February 28th are utilized to perform simulation regarding robust and on-line CD scheduling. Note that, although the initial SVR model is trained utilizing demand data from Jan. 1st to Jan. 20th (i.e. 20 days), the SVR model is kept updated using the immediate past 20-days demand data (from the simulation day). For example, while performing simulation for Jan. 25th, the demand data from Jan. 5th to Jan. 24th are utilized.
\end{itemize}

The algorithms are implemented in \emph{Python} programming language. The Bayesian MCMC based group formation algorithm is implemented in conjunction with PyMC package \citep{patil2010pymc}, an open-source \emph{Python} package to perform Bayesian analysis. The open-source solver \citep{cbc} is used to solve the MILP in robust CD scheduling.

\subsection{Formation of Balancing Groups}
The 1st part of this section describes the analysis and results regarding energy balancing group formation. The demand data has a 30-minutes granularity. The $MDSD$ (i.e. $DAC$) of each customer are determined according to Eq. \ref{eqn_mdsd} considering demands from January 1st, 2013 to January 20th, 2013.

\begin{figure}[htb]
\centering
\includegraphics[scale=0.50]{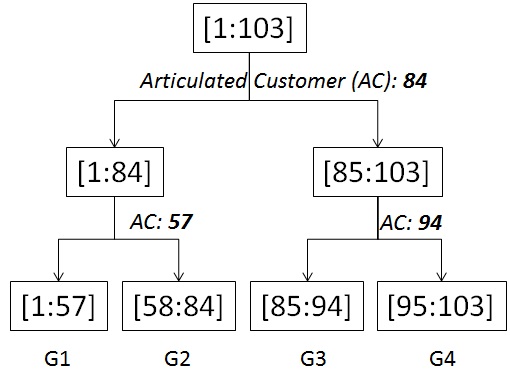}
\caption{Divide-and-conquer process of group formation: Status of \emph{dacVector}.}
\label{fig_dnc_group}
\end{figure}

Figure~\ref{fig_dnc_group}: \textit{Divide-and-conquer process of group formation: Status of \emph{dacVector}.}

\begin{figure}[htb]
\centering
\includegraphics[scale=0.55]{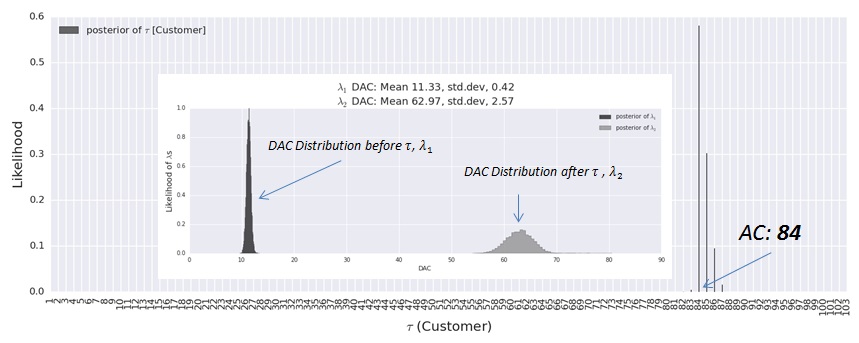}
\caption{The posterior distributions of $\tau$ and $\lambda$s after performing Bayesian MCMC at 1st Phase .}
\label{fig_dac_dist}
\end{figure}

The divide-and-conquer process of group formation with associated $dacVector$ (Figure~\ref{fig_demand_strategy}) range is shown in Figure~\ref{fig_dnc_group}. The corresponding articulated customers at each stage of division are also pointed out in the figure. For example, initially the $dacVector$ contains customers from 1 to 103, $[1,103]$. After performing the Bayesian MCMC sampling method on $dacVector$, customer 84 is selected as articulated customer (AC) and hence serves as the dividing point (1st Phase). The posterior distributions of the model parameters $\tau$ and $\lambda$s at 1st Phase are plotted in Figure~\ref{fig_dac_dist}. The distributions of $\tau$ identifies customer 84 as AC since it is highly likely to be one. Note that, the prior of $\tau$ was a uniform distribution, which is changed after performing the Bayesian MCMC sampling process and provides a posterior that correctly maps to the target true distribution (i.e. \emph{observation}).
The distributions of DACs (i.e. $\lambda$s, before and after $\tau$, respectively) are shown in the figure as well. Although, the $\lambda$s were set as \emph{Exponential Distributions} as priors, the posteriors come out as \emph{Normal Distributions} through the Bayesian inference. In the process of MCMC, a total number of 80,000 random samples are generated utilizing the prior distributions of model parameters. Among them 25\% of the samples are discarded (so called \emph{burn-in} \citep{geyer2011introduction} of samples since the convergence of the \emph{Markov Chain} is not fully known) while tracing the posterior samples.
Referring back to Figure~\ref{fig_dnc_group}, at the next phase, the $dacVector$ is divided into two child vectors of $[1,84]$ and $[85,103]$, each of which will go through the Bayesian MCMC process.The process is repeated until the vector is not further divisible (i.e. $\Delta_{dac}$ is below a point, $T$), and thereby forming balancing groups. Finally, the process settles down forming 4 balancing groups. 

\begin{figure}[htb]
\centering
\includegraphics[scale=0.4]{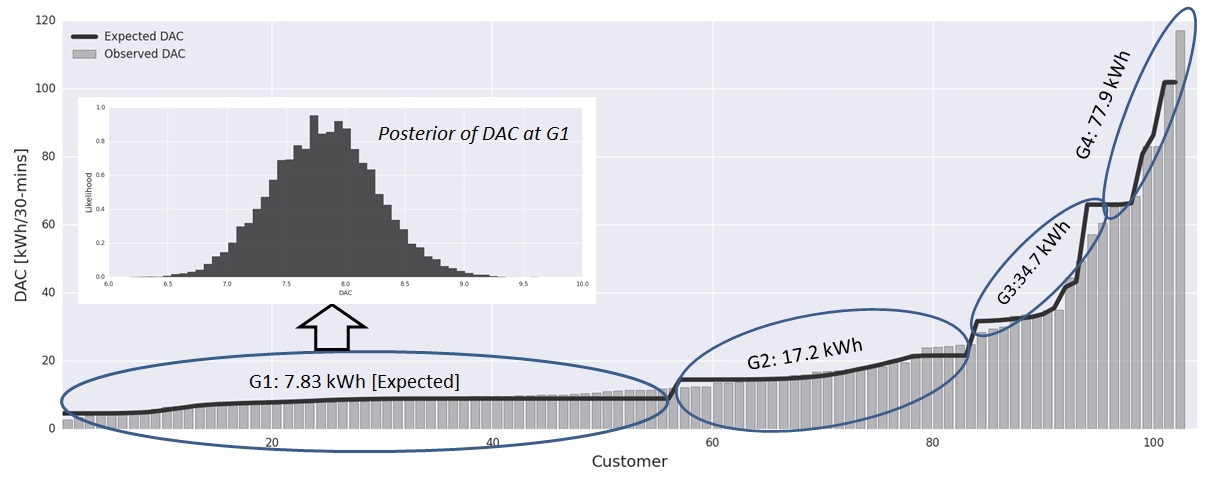}
\caption{Group formation using Bayesian MCMC based aggregation strategy.}
\label{fig_grp_formation}
\end{figure}

\begin{figure}[htb]
\centering
\includegraphics[scale=0.60]{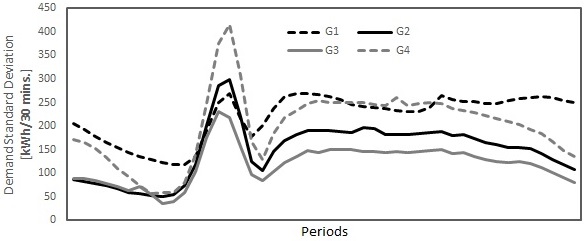}
\caption{Resultant (periodical) demand standard deviation for each group.}
\label{fig_grp_dsd}
\end{figure}

Figure~\ref{fig_grp_formation} shows the result of Bayesian MCMC based demand aggregation and resultant balancing groups (the customers are already ordered according to their $DSD$). At the same time, the figure points out the expected $DSD$ per customer in a group. For example, Group 1 (G1), that contains 57 customers, has an expected $DSD$ of 7.83kWh per customer. As a by-product of Bayesian MCMC method, the posterior distribution of $DSD$ over the customers in G1 is evaluated, which is also printed in Figure~\ref{fig_grp_formation} as a \emph{Normal Distribution}. The periodical (half-hourly) aggregated $DSD$ for each group is shown in Figure~\ref{fig_grp_dsd}. Evidently, the Bayesian MCMC based demand aggregation strategy forms groups whose $DSD$s are close to each other even when the group sizes are different. That establishes the proposed demand aggregation strategy is an effective clustering strategy of relatively heterogeneous demand signals.

Figure~\ref{fig_dac_dist}: \textit{The posterior distributions of $\tau$ and $\lambda$s after performing Bayesian MCMC at 1st Phase .}

Figure~\ref{fig_grp_formation}: \textit{Group formation using Bayesian MCMC based aggregation strategy.}

Figure~\ref{fig_grp_dsd}: \textit{Resultant (periodical) demand standard deviation for each group.}

\subsection{Robust CD Scheduling of Energy Storage}
At the 2nd part, we will investigate the on-line robust energy storage scheduling (S-SWCD) to reduce imbalance energy and cost for Group 1 (G1). The G1 contains 57 customers. The first 20 days' (January 1 to 20) demand data is utilized to train the initial demand predictor model, which gives us 38 days (approximately) to perform the on-line CD scheduling algorithm and evaluate the performance. The peak demand (kWh/30-minutes) for the first 20-days is recorded as 2,500kWh/30-minutes.

In order to perform the short-term demand prediction, prediction window size, $w$ is set as 8. {The initial \emph{scenario space}, $\mathcal{S}$ contains 5000 scenarios. Finally, the number of scenarios in the reduced scenario set, $S$ is set as 57. The values of $w$ and $|S|$ are settled to 8 and 57, respectively based on the sensitivity analysis of these parameters on the imbalance cost reduction while keeping the battery capacity fixed.}
As an energy storage, Lithium-ion Battery is utilized. The size of the battery (power rating and consequently energy capacity) are determined by analyzing the $DAC$ distribution (e.g. as in Figure~\ref{fig_grp_formation}), since $DAC$ is essentially $MDSD$ which in turn represents the peak potential deviation that requires to be minimized by battery.
The expected $DAC$ for G1, 7.83kWh/customer, can be utilized as battery size (that makes an approximated 445kWh of aggregated battery capacity for G1). However, due to higher accuracy in day-ahead prediction, the aggregated $DAC$ (i.e. potential battery capacity) of 445kWh can be treated as an upper-bound. In the experiment, we present the results considering aggregated battery capacity of 320kWh. At the same time, the power rating of the aggregated batteries is considered as 640 kW (i.e. the batteries are of 2E ratings). A total of 50 batteries (of rating 6.4kWh/12.8kW, SOC limit of 1\% to 96\% and C/D round-trip efficiency of 95\%) are assumed to be installed at G1. These batteries are operated and controlled (via PPS's SCADA system) synchronously. Therefore, the simulations are performed considering aggregated battery capacity and aggregated SOC. We assume a day-ahead prediction error (i.e. the basic imbalance due to deviation between contracted supply and actual demand) Normally Distributed around 0 mean and having 10\% (of demand) as a standard deviation. The imbalance energy threshold beyond which the penalty tariff is applied is set as 93kWh (3\% of the maximum contracted supply)\citep{eim}.

\begin{figure}[htb]
\centering
\includegraphics[scale=0.6]{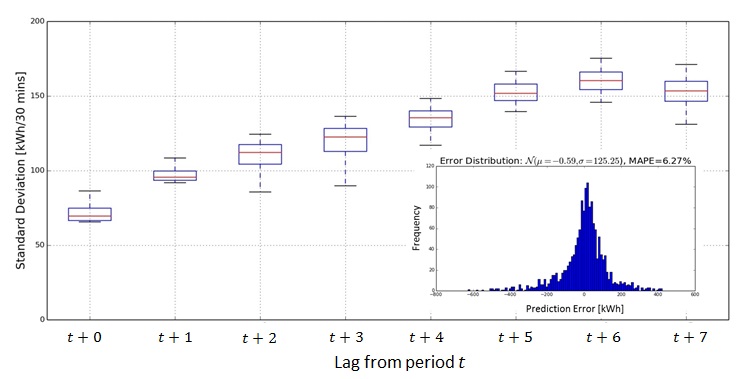}
\caption{Prediction error statistics (lag-based) with combined error distribution.}
\label{fig_prd_error}
\end{figure}

The performance of SVR based short-term demand predictor is described in Figure~\ref{fig_prd_error}. The lag-wise error distributions (standard deviations) follow a pattern where the errors are relatively lower in smaller lags. The accumulated error distribution over all lags are depicted in the lower right corner of the figure. The errors are normally distributed ($\mathcal{N}(-0.59, 125.25kWh)$) with a Mean Absolute Percent Error (MAPE) of 6.27\%. The prediction accuracy is therefore good enough to be utilized at the on-line optimization. In order to generate scenarios, the lag-based error statistics (mean and standard deviation, as of Eq. \ref{eqn_scn_gen}) is utilized. The generated scenarios are referred back to Figure~\ref{fig_scn_gen}.

Figure~\ref{fig_prd_error}: \textit{Prediction error statistics (lag-based) with combined error distribution.}

\begin{figure}[htb]
\centering
\includegraphics[scale=0.70]{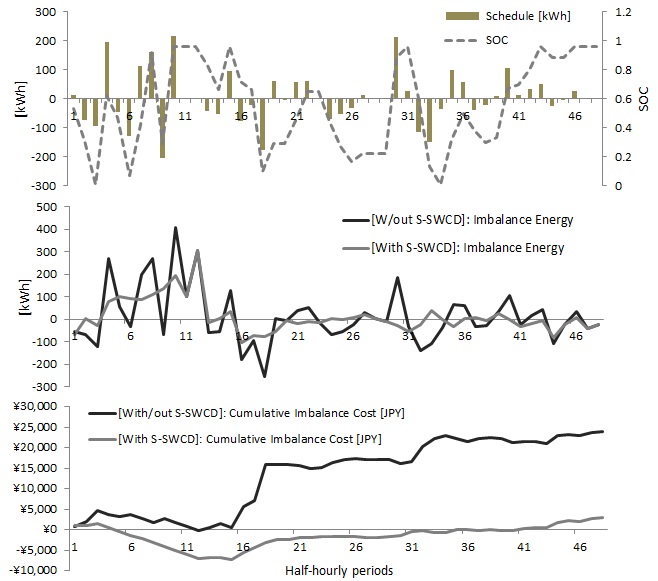}
\caption{Battery CD scheduling with S-SWCD and resultant imbalance energy pattern and cost comparison [1-day].}
\label{fig_storage_schedule}
\end{figure}

Figure~\ref{fig_storage_schedule}: \textit{Battery CD scheduling with S-SWCD and resultant imbalance energy pattern and cost comparison [1-day].}

\begin{figure}[htb]
\centering
\includegraphics[scale=0.5]{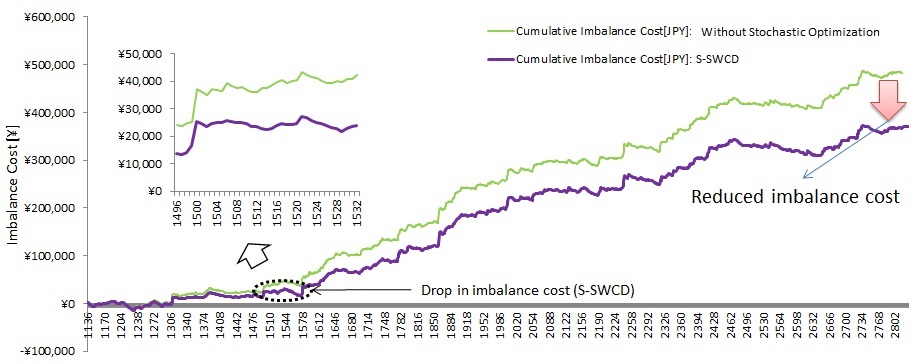}
\caption{Cumulative imbalance cost reduction pattern using S-SWCD with a deterministic equivalent algorithm [accumulated over 38-days].}
\label{fig_cumulative_ic}
\end{figure}

Figure~\ref{fig_cumulative_ic}: \textit{Cumulative imbalance cost reduction pattern using S-SWCD with a deterministic equivalent algorithm [accumulated over 38-days].}

The half-hourly battery scheduling, imbalance energy reduction, and imbalance cost reduction patterns (operated for 1-day) utilizing S-SWCD method (with aggregated battery capacity of 320kWh) are plotted in Figure~\ref{fig_storage_schedule}. The battery CD schedule with power dispatch (-ve for discharge and +ve for charging) and associated SOC is drawn in the top figure. The resultant imbalance (energy) reduction is shown in middle figure. Evidently, the S-SWCD tries to keep the imbalance energy within the limit of imbalance threshold (i.e. $\pm93$kWh), if not zero, by intelligently scheduling and dispatching the energy from/to the batteries. The consequent (cumulative) imbalance cost reduction pattern (compared to basic imbalance cost without deploying batteries plus S-SWCD) is plotted as the bottom figure. Clearly, the imbalance cost is significantly reduced by applying batteries with S-SWCD controlling algorithm. Note that, at certain period, the cumulative cost goes down because, at that period, PPS makes revenue by selling energy back to EIM (as the supply is higher than the demand). 

The 38-days imbalance cost reduction pattern is shown in Figure~\ref{fig_cumulative_ic}. In order to compare the performance of S-SWCD, we have implemented an equivalent intelligent algorithm that utilizes the deterministic (non-stochastic) predicted demand signal (i.e. considering only 1 scenario, which is the predicted signal) and applied the same amount of battery storages (320kWh). As seen in the Figure~\ref{fig_cumulative_ic}, S-SWCD outperforms its non-stochastic counter part by a good margin (almost 23\% reduction in imbalance cost after the end of 38-days). Initially, both of these algorithms attain approximately similar imbalance cost reduction. However, from the periods 1500 onwards, a significant jump in cumulative imbalance cost is experienced by the deterministic algorithm, where the jump in corresponding S-SWCD is relatively lower. Such improvement in S-SWCD is realized by better planning and scheduling of battery storage so that in the critical moments (period 1500 and such) the battery storage is able to deliver appropriate energy that avoids costly EIM interactions. The basic imbalance cost (without battery) for 38-days is reported as 1,852,266 JPY where the cost incurred via S-SWCD is approximately 372,860 JPY (i.e. an 80\% reduction in imbalance cost). 

\begin{figure}[htb]
\centering
\includegraphics[scale=0.5]{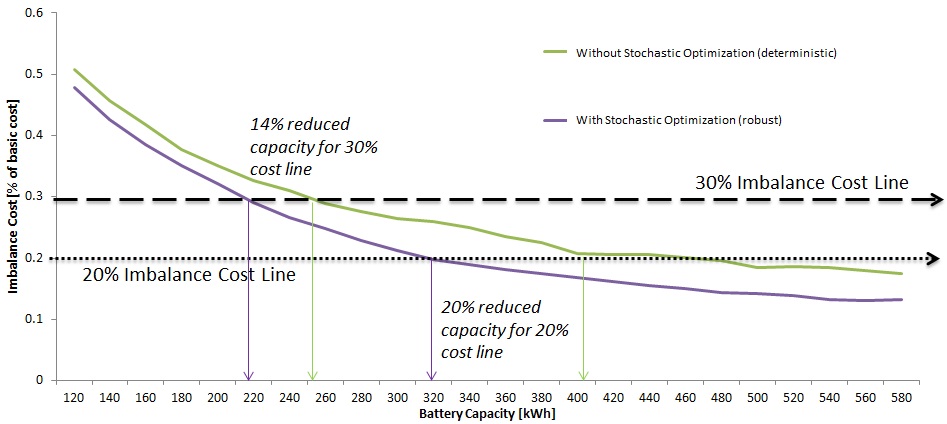}
\caption{Imbalance cost reduction pattern (via S-SWCD) with increasing battery capacity.}
\label{fig_cost_vs_capacity}
\end{figure}

Figure~\ref{fig_cost_vs_capacity}: \textit{Imbalance cost reduction pattern (via S-SWCD) with increasing battery capacity.}

The cost reduction, however, depends on the size of battery storage. To this end, Figure~\ref{fig_cost_vs_capacity} is presented to show the effect of storage capacity with imbalance cost reduction (as a percentile of basic imbalance cost incurred without any storage). The figure points interesting insights such as, no matter how large the aggregated capacity is, full reduction of imbalance cost is never possible. Moreover, imbalance cost tends to decrease faster with relatively lower battery capacity. A trade-off therefore exists between battery cost and imbalance cost where PPS needs to decide the target line of imbalance cost without investing too much on battery. A target line (of 20\% of basic imbalance cost) is drawn to notice the required battery capacity to reach the line. The equivalent curve generated by the deterministic scheduler is plotted to compare with the S-SWCD. S-SWCD requires 320kWh of storage while the deterministic scheduler requires 400kWh (i.e. S-SWCD requires 20\% less battery capacity to reach a target of 20\% of basic imbalance cost). Depending on the target line settings, the performance of S-SWCD (as far as storage capacity is concerned) varies. For example, S-SWCD takes 14\% less battery (compared to the deterministic scheduler) when the target line is set as 30\% of the basic imbalance cost.

Finally, it is worthy to mention that, the energy balancing group formation effectively provides stable grouping, since the $MDSD$ of a customer does not change significantly. Therefore, the group formation can be done in off-line planning stage, which is perfect for a day- or week-ahead energy procurement process. However, in case of significant change in a customer's $MDSD$ (that results group reformation), certain re-planning might require.

\section{Conclusion}
\label{conclusion}
In this paper, we propose novel solution strategies for two of the fundamental yet interconnected problems of PPS. The 1st problem deals with formation of energy based balancing groups within a particular class of customers. In this paper, apartment buildings are taken as customers. To solve the group formation problem, we apply a Bayesian MCMC based divide-and-conquer strategy that takes a statistical measure (periodical demand standard deviation of each customer) and produces a number of balancing groups. The purpose of creating balancing group is to choose similar customers whose aggregated demand has higher predictability, which helps PPS to plan the supply contract. Bayesian MCMC is proven to be an effective aggregation strategy that not only identifies appropriate grouping but also provides important demand centric insights regarding each group (e.g. the upper-bound of required storage capacity for imbalance reduction). However, due to the uncertainty in actual demand, the imbalance occurrence is inevitable. This brings us to the 2nd problem of on-line imbalance energy and cost reduction on the face of demand uncertainty. We propose a robust on-line CD scheduling for energy storage (batteries) that reduces the expected imbalance energy and cost (solving a multi-objective optimization problem) by incorporating the non-linearity in imbalance tariff, storage dynamics and associated non-linear constraints. The experimental results prove that the proposed scheduling method is robust against demand prediction uncertainty and is capable of lowering imbalance cost with minimized storage capacity. At the same time, we provide insights regarding the trade-off between imbalance reduction and storage capacity which tend to stimulate the investment related issue for a PPS. Although, the problems addressed in this paper focus on Japanese power market, the solutions can be effectively utilized in other markets as well by making appropriate assumptions. For example, in Europe, Balance Responsible Parties (BRPs) are market participants, consisting of controllable and uncontrollable generators
and loads, that are legally entitled to trade electricity on the various power markets (forward, day-ahead, ancillary services, intra-day, imbalance markets) in order to satisfy loads within their control area, earn profit and contribute to
the preservation of power balance in the power grid. Therefore, the proposed methodology can be plugged-in into the European energy market with minimal changes that align with regularity and policy. Possible future research avenue will be analyzing the effects of supply uncertainty as well as dynamic imbalance tariff.

\section*{Acknowledgement}
The authors would like to thank Digital Grid Co., Ltd. and the University of Tokyo. This work is a part of collaborative research program with these entities.

\newpage

\begin{thebibliography}{10}
\expandafter\ifx\csname url\endcsname\relax
  \def\url#1{\texttt{#1}}\fi
\expandafter\ifx\csname urlprefix\endcsname\relax\def\urlprefix{URL }\fi
\expandafter\ifx\csname href\endcsname\relax
  \def\href#1#2{#2} \def\path#1{#1}\fi

\bibitem{meti}
A.~N. Resources, Energy,
  \href{http://www.meti.go.jp/english/policy/energy\_environment\\/electricity\_system\_reform/pdf/201410EMR\_in\_Japan.pdf}{Electricity
  market reform in japan}, Tech. rep., Ministry of Economy, Trade and Industry
  (Oct 2014).
\newline\urlprefix\url{http://www.meti.go.jp/english/policy/energy\_environment\\/electricity\_system\_reform/pdf/201410EMR\_in\_Japan.pdf}

\bibitem{Abe:2011}
R.~Abe, H.~Taoka, D.~McQuilkin, Digital grid: Communicative electrical grids of
  the future, Smart Grid, IEEE Transactions on 2~(2) (2011) 399--410.
\newblock \href {http://dx.doi.org/10.1109/TSG.2011.2132744}
  {\path{doi:10.1109/TSG.2011.2132744}}.

\bibitem{eim}
P.~Study-group,
  \href{http://ieei.or.jp/wp-content/uploads/2013/04/8e4776f6f89c3d5cb1d531d3f7fe6e23.pdf}{Electric
  power industry reform}, Tech. rep., International Environment and Economy
  Institute (Apr 2013).
\newline\urlprefix\url{http://ieei.or.jp/wp-content/uploads/2013/04/8e4776f6f89c3d5cb1d531d3f7fe6e23.pdf}

\bibitem{gordon2014probabilistic}
A.~D. Gordon, T.~A. Henzinger, A.~V. Nori, S.~K. Rajamani, Probabilistic
  programming, in: Proceedings of the on Future of Software Engineering, ACM,
  2014, pp. 167--181.

\bibitem{geyer2011introduction}
C.~Geyer, Introduction to markov chain monte carlo, Handbook of Markov Chain
  Monte Carlo (2011) 3--48.

\bibitem{gelman2014bayesian}
A.~Gelman, J.~B. Carlin, H.~S. Stern, D.~B. Rubin, Bayesian data analysis,
  Vol.~2, Taylor \& Francis, 2014.

\bibitem{Boyd:2004:CO:993483}
S.~Boyd, L.~Vandenberghe, Convex Optimization, Cambridge University Press, New
  York, NY, USA, 2004.

\bibitem{Bemporad1999407}
A.~Bemporad, M.~Morari, Control of systems integrating logic, dynamics, and
  constraints, Automatica 35~(3) (1999) 407 -- 427.

\bibitem{cbc}
P.~Coin-OR, \href{http://projects.coin-or.org/Cbc}{Cbc, coin-or branch and cut}
  (2010).
\newline\urlprefix\url{http://projects.coin-or.org/Cbc}

\bibitem{Dias20147722}
J.~G. Dias, S.~B. Ramos, Dynamic clustering of energy markets: An extended
  hidden markov approach, Expert Systems with Applications 41~(17) (2014) 7722
  -- 7729.

\bibitem{Chicco201268}
G.~Chicco, Overview and performance assessment of the clustering methods for
  electrical load pattern grouping, Energy 42~(1) (2012) 68 -- 80.

\bibitem{1626400}
G.~Chicco, R.~Napoli, F.~Piglione, Comparisons among clustering techniques for
  electricity customer classification, IEEE Transactions on Power Systems
  21~(2) (2006) 933--940.

\bibitem{7063233}
S.~Haben, C.~Singleton, P.~Grindrod, Analysis and clustering of residential
  customers energy behavioral demand using smart meter data, Smart Grid, IEEE
  Transactions on 7~(1) (2016) 136--144.

\bibitem{firestone2005energy}
R.~Firestone, C.~Marnay, Energy manager design for microgrids, Lawrence
  Berkeley National Laboratory.

\bibitem{6417004}
R.~Palma-Behnke, C.~Benavides, F.~Lanas, B.~Severino, L.~Reyes, J.~Llanos,
  D.~Saez, A microgrid energy management system based on the rolling horizon
  strategy, Smart Grid, IEEE Transactions on 4~(2) (2013) 996--1006.

\bibitem{5587912}
P.~Meibom, R.~Barth, B.~Hasche, H.~Brand, C.~Weber, M.~O'Malley, Stochastic
  optimization model to study the operational impacts of high wind penetrations
  in ireland, Power Systems, IEEE Transactions on 26~(3) (2011) 1367--1379.

\bibitem{Wang2015597}
R.~Wang, P.~Wang, G.~Xiao, A robust optimization approach for energy generation
  scheduling in microgrids, Energy Conversion and Management 106 (2015) 597 --
  607.

\bibitem{6802411}
Y.~Ma, J.~Matuško, F.~Borrelli, Stochastic model predictive control for
  building hvac systems: Complexity and conservatism, IEEE Transactions on
  Control Systems Technology 23~(1) (2015) 101--116.

\bibitem{viana2013new}
A.~Viana, J.~P. Pedroso, A new milp-based approach for unit commitment in power
  production planning, International Journal of Electrical Power \& Energy
  Systems 44~(1) (2013) 997--1005.

\bibitem{1664974}
M.~Carrion, J.~M. Arroyo, A computationally efficient mixed-integer linear
  formulation for the thermal unit commitment problem, IEEE Transactions on
  Power Systems 21~(3) (2006) 1371--1378.

\bibitem{burges1998tutorial}
C.~J. Burges, A tutorial on support vector machines for pattern recognition,
  Data mining and knowledge discovery 2~(2) (1998) 121--167.

\bibitem{hoyland2001generating}
K.~H{\o}yland, S.~W. Wallace, Generating scenario trees for multistage decision
  problems, Management Science 47~(2) (2001) 295--307.

\bibitem{patil2010pymc}
A.~Patil, D.~Huard, C.~J. Fonnesbeck, Pymc: Bayesian stochastic modelling in
  python, Journal of statistical software 35~(4) (2010) 1.

\end{thebibliography}
\end{document}